\documentclass[reprint,amsmath,amssymb,aps]{revtex4-2}

\usepackage{moresize}
\usepackage{graphicx}
\usepackage{amsmath}
\usepackage{booktabs}
\usepackage{xcolor}
\usepackage{dcolumn}
\usepackage{bm}

\begin{document}

\preprint{APS/123-QED}

\title{Hybrid Ferromagnet-SNSPDs: Single photon induced order-to-disorder transition in ferromagnets coupled to thin film superconductors} 

\author{Leif Bauer}\altaffiliation{These authors contributed equally to the work.}\affiliation{The Elmore Family School of Electrical and Computer Engineering, Purdue University, West Lafayette, 47907, IN, USA}
\author{Daien He}\altaffiliation{These authors contributed equally to the work.}\affiliation{The Elmore Family School of Electrical and Computer Engineering, Purdue University, West Lafayette, 47907, IN, USA}
\author{Sathwik Bharadwaj}
\affiliation{The Elmore Family School of Electrical and Computer Engineering, Purdue University, West Lafayette, 47907, IN, USA}
\affiliation{Department of Physics, Worcester Polytechnic Institute, Worcester, 01609, MA, USA}
\author{Zubin Jacob}\email{Contact author: zjacob@purdue.edu}
\affiliation{The Elmore Family School of Electrical and Computer Engineering, Purdue University, West Lafayette, 47907, IN, USA}

\date{\today}

\begin{abstract}
The development of midwave and longwave infrared single photon detectors is crucial for their emerging applications in spectroscopy, remote sensing, exoplanet detection, and free space quantum communications. However, existing sensors need to be operated at extremely low temperatures (0.08-0.9K) to reduce dark noise and hence require the use of advanced cryogenics such as dilution refrigerators or $^3$He cryogens, significantly limiting applications. Here we propose a vortex-engineering approach based on a hybrid phase transition in a ferromagnet/superconductor bilayer to increase the operating temperature of infrared single photon detectors up to 3.75K. We show that the introduction of a  ferromagnetic layer produces a local magnetic field which impedes vortex crossing in the superconductor, reducing dark noise. When a single photon is incident, the photon-induced hotspot causes an order-to-disorder transition in the ferromagnet, leading to a vortex-induced phase transition in the superconducting layer. By engineering the ferromagnet's Curie temperature to be close to the device's operating temperature, single photon sensitivity can be achieved at increased operating temperatures. We predict at midwave/longwave infrared wavelengths (3-14$\mu$m) the operating temperature can be raised to 3.25-3.75K, enabling significantly simpler cooling systems. 
\end{abstract}

\maketitle 
\section{Introduction}
\begin{figure*}[ht!]
\centering
\includegraphics[width=\textwidth]{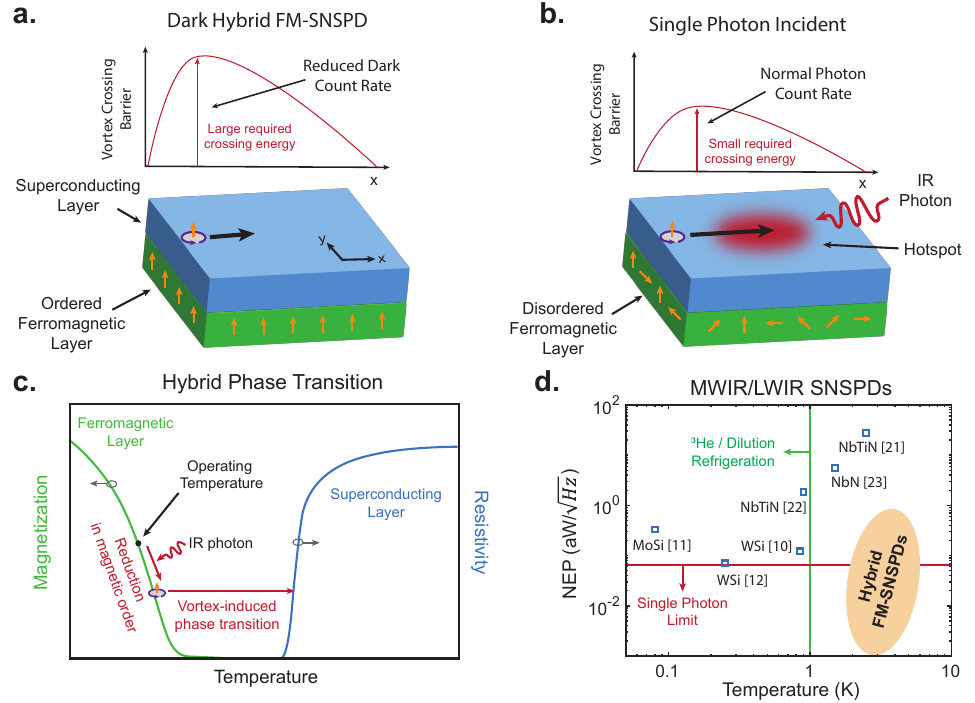}
\caption{Hybrid phase transition in a ferromagnet/superconductor (FM-S) bilayer. (a) When no photons are incident the vortex crossing barrier is large reducing dark counts. This large barrier is due to the presence of a magnetic layer with magnetization (yellow arrows) aligned to the vortex magnetic flux (yellow arrow). (b) When a single photon is incident, the vortex barrier is reduced enabling high photon detection efficiency. The barrier is reduced after photon incidence due to a reduction in the local order of the magnetic layer from the photon-induced hot spot. This reduced ferromagnetic order, causes a reduction in the magnetic field in the superconducting layer, leading to a reduced vortex barrier. (c) Hybrid phase transition of ferromagnet and superconductor in FM-S bilayer. We choose a ferromagnet with Curie temperature close to the operating temperature such that a small temperature increase leads to a large change in magnetization. A hotspot from an incident single photon causes a reduction in the ferromagnetic order, reducing the magnetic field in the superconducting layer. The reduced field combined with the hotspot in the superconducting layer causes a single vortex crossing event to occur, leading to a phase transition in the superconductor to the normal metal state. (d) Noise equivalent power (NEP) comparison of hybrid FM-SNSPD (this work) versus SNSPD experiments \cite{verma_single-photon_2021,chen_mid-infrared_2021,taylor_low-noise_2023,chang_efficient_2022,azem_mid-infrared_2024,marsili_efficient_2012} in the MWIR/LWIR (lower is more sensitive). NEP below the marked single photon limit indicates SNR $>$ 1 for incident photon flux of 1 photon/s at 1Hz readout bandwidth. We predict hybrid FM-SNSPDs will have reduced noise, enabling LWIR single photon sensitivity without the use of advanced cryocooling setups.}
\label{Fig:M_SNSPD}
\vspace{-10pt}
\end{figure*}

The development of single photon detectors in the midwave and longwave infrared  (3-14$\mu$m) has recently generated considerable interest due to their potential applications in infrared spectroscopy \cite{lau_superconducting_2023}, quantum remote sensing \cite{bao_photon_2024}, dark matter searches \cite{baudis_first_2025}, exoplanet detection \cite{wollman_recent_2021}, and free space quantum communication \cite{dello_russo_advances_2022}. In particular, superconducting nanowire single photon detectors (SNSPDs) are an attractive option due to their low reset times \cite{cherednichenko_low_2021}, high timing resolution \cite{korzh_demonstration_2020} and high quantum efficiencies \cite{reddy_superconducting_2020}. Currently, midwave and longwave infrared SNSPDs rely on materials with small superconducting energy gaps to enable high detection efficiency at these longer wavelengths \cite{colangelo_large-area_2022,verma_single-photon_2021,chen_mid-infrared_2021,taylor_low-noise_2023}. This small superconducting energy gap also causes an increase in thermally-activated dark counts \cite{jahani_probabilistic_2020}, requiring lower operating temperatures ($T<1$K) to enable single photon sensitivity (see Fig. \ref{Fig:M_SNSPD}d). These operating temperatures require advanced cooling solutions such as dilution refrigerators or $^3$He cryogens, leading to significant size, cost, and power requirements. Here we propose a ferromagnet/superconductor bilayer designed to achieve significantly reduced dark counts at higher operating temperatures (T$\sim$3-4 K) for mid- and long-wave infrared single photon detectors.  We identify a mechanism in which the incident single photon induces an order-to-disorder phase transition in the ferromagnet, which further leads to a vortex-induced phase transition in the superconducting layer (see Fig. \ref{Fig:M_SNSPD}c). 

The influence of magnetic fields on superconductors has been well studied, with publications dating back nearly 70 years \cite{abrikosov_magnetic_1957}. Large applied magnetic fields cause type-2 superconductors to form a vortex lattice where vortex density increases proportional to the magnetic field until the superconductor breaks down  \cite{tinkham_introduction_2015}. Small applied magnetic fields do not produce a vortex lattice \cite{tinkham_introduction_2015}, but can increase the vortex crossing barrier \cite{bulaevskii_vortex-assisted_2012}, leading to fewer vortex crossing events/dark counts \cite{jahani_probabilistic_2020}. Several SNSPD experiments have demonstrated this reduction in dark counts, which depends on the direction of the magnetic field \cite{charaev_magnetic-field_2019,lawrie_multifunctional_2021,engel_dependence_2012}. This dark count reduction has also been experimentally demonstrated in ferromagnet/superconductor bilayers, where the addition of a ferromagnetic layer results in a reduction in the dark count rate \cite{cristiano_superconductorferromagnet_2016}. However, the large barrier which reduces dark counts will also limit the detection efficiency of photon-induced counts, limiting the detector sensitivity \cite{jahani_probabilistic_2020}.

An ideal device should have a small magnetic field when dark, and a significantly reduced magnetic field upon photon incidence. Here we utilize a ferromagnetic layer which undergoes an order-disorder transition after photon incidence. As shown in Fig. \ref{Fig:M_SNSPD}a,b, a small magnetic field is produced by the spontaneous magnetization of the ferromagnetic layer, and after photon incidence the ferromagnet becomes disordered, leading to a reduced magnetic field. In this design, the response of the ferromagnetic layer is intimately linked to the vortex dynamics in the superconducting layer. We require the ferromagnet to produce a strong enough magnetic field ($\sim$10-40mT) \cite{charaev_magnetic-field_2019} to significantly reduce the dark count rate. Simultaneously, we require a large reduction in magnetization as the temperature increases, causing the magnetic field to significantly decrease after photon incidence. We achieve this by selecting or engineering a ferromagnet with a Curie temperature close to the device operating temperature with a thickness large enough to produce significant magnetic fields in the superconducting layer.

In this paper, we analyze the effect of a hybrid ferromagnet-superconductor phase transition on the sensitivity of infrared single photon detectors. We propose several potential ferromagnet/superconductor combinations for different operating temperatures and wavelength ranges. Utilizing a vortex-crossing theory of detection, we demonstrate an increase in sensitivity in these hybrid devices. This will result in larger operating temperatures ($T>1K$) across a range of wavelengths from visible to LWIR. Lastly, we present atomistic spin dynamics simulations of a potential synthetic ferromagnetic layer based on a Ni$_x$Cu$_{1-x}$ alloy. Using these simulations we find the magnetic anisotropy and Curie temperature of Ni$_x$Cu$_{1-x}$, and identify concentrations with a low Curie temperature. We demonstrate that this engineered ferromagnetic layer is capable of producing significant magnetic fields in the superconducting layer, providing a pathway to ferromagnet-based vortex engineering.

\section{Theory of Hybrid Ferromagnet-Superconductor Phase Transitions in SNSPDs}

We employ the single vortex model of SNSPD detection to model device metrics, which has recently been demonstrated to be effective in modeling dark count rate, system detection efficiency, and timing jitter \cite{engel_detection_2015,he_unified_2025}. In this model an incident photon is absorbed in the superconductor creating a hotspot which lowers the vortex crossing barrier leading to a vortex crossing event \cite{jahani_probabilistic_2020}. As the vortex crosses the superconductor, the superconducting phase is disturbed, causing a 2$\pi$ phase shift across the vortex path \cite{PhysRevB.77.174517}. This phase slip causes a phase transition to the normal metal state, i.e. a `count' \cite{PhysRevB.77.174517}. The single vortex model also describes dark count behavior, where latent thermal energy causes vortex crossing events to occur at a rate depending on the size of the vortex crossing barrier and the device temperature \cite{PhysRevB.83.144526,he_unified_2025}.

Traditionally, infrared SNSPDs utilize temperature and current biasing to achieve single photon sensitivity with low levels of dark counts \cite{engel_detection_2015,he_unified_2025,jahani_probabilistic_2020,verma_single-photon_2021,chen_mid-infrared_2021,taylor_low-noise_2023,chang_efficient_2022,marsili_efficient_2012,azem_mid-infrared_2024}. We propose a different class of SNSPDs which utilize ferromagnetic and superconducting phase transitions which we call hybrid FM-SNSPDs. The design consists of a ferromagnetic layer below a superconducting layer in a traditional nanowire pattern (see Fig. \ref{Fig:M_SNSPD}(a)). The purpose of the ferromagnetic layer is to provide a magnetic field in the superconducting layer which increases the vortex crossing barrier. The barrier in the presence of a magnetic field is given by \cite{engel_detection_2015,bulaevskii_vortex-assisted_2012}
\begin{eqnarray}
    \frac{U_{max}}{\epsilon_0} =\nonumber && \max_{x_v}\left[\ln{\left(\frac{2W}{\pi\xi}\sin{(\pi x_v)}\right)}\right.\\&&\left. - \frac{I}{I_c}\frac{2x_v}{\exp(1)\xi} + \frac{H}{H_0}x_v\left(1-\frac{x_v}{\pi}\right)\right]
    \label{eq:potential_function}
\end{eqnarray}
\begin{equation}
    \epsilon_0 = \frac{\Phi_0^2d}{4\pi\mu_0\lambda^2}
\end{equation}
where $U_{max}$ is the vortex crossing barrier, $x_\nu = x/W$ is the normalized vortex position, $W$ is the nanowire width, $I$ is the bias current, $I_c$ is the critical current, $\xi$ is the coherence length, $H$ is the applied magnetic field, $H_0 = \Phi_0/2W^2$, $\Phi_0$ is the magnetic flux quantum, $\epsilon_0$ is the vortex energy, $d$ is the superconducting nanowire thickness, $\mu_0$ is the vacuum permeability, and $\lambda$ is the magnetic penetration depth.

The system detection efficiency and dark count rate are both affected by the size of the vortex crossing barrier $U_{max}$. Following thermal activation theory \cite{hanggi_reaction-rate_1990}, the vortex crossing rate is given by 
\begin{equation}
    \Gamma_v = \alpha_v\exp(-U_{max}/k_BT)
    \label{eq:DCR}
\end{equation}
where $\alpha_v$ is the vortex attempt rate (a material dependent fitting parameter) and $\Gamma_v$ is the vortex crossing rate. The dark count rate is equal to the vortex crossing rate when no photons are incident
\begin{equation}
    DCR = \Gamma_v.
\end{equation}
After a photon is absorbed, a hot spot is created in the superconductor which causes a quasiparticle diffusion process that changes the current density near the hot spot \cite{engel_detection_2015}. This results in a time-varying change in the vortex crossing barrier (see the Supplementary Material for further details \cite{supp}) \cite{engel_detection_2015,engel_numerical_2013,jahani_probabilistic_2020}. The system detection efficiency (SDE) is obtained from the time varying change in vortex crossing rate using the following equation \cite{jahani_probabilistic_2020,he_unified_2025}
\begin{equation}
    SDE=1-\exp\left(-\int^t_0\Gamma_{v}(t')dt'\right).
    \label{eq:SDE}
\end{equation}

The relative direction of the magnetic field and current in Eq. \ref{eq:potential_function} are important since they determine whether the device has a larger or smaller count rate. We will choose the field to be parallel to the orientation of the vortex's magnetic flux (see Fig. \ref{Fig:M_SNSPD}(a)) such that the hybrid FM-SNSPD produces a smaller dark count rate. This design results in a larger barrier that reduces SNSPD noise but also reduces sensitivity to single photon events. To maintain single photon sensitivity, we also require the ferromagnetic layer to have a Curie temperature close to the operating temperature of the SNSPD. Thus the ferromagnet design requirements are (1) a perpendicular magnetic anisotropy, (2) a Curie temperature near the operating temperature, and (3) a thermal time constant longer than the photon-induced changes in the vortex crossing barrier. This design enables a hybrid phase transition of the device upon photon incidence, where the ferromagnetic layer will experience a local temperature increase due to photon absorption (see Fig. \ref{Fig:M_SNSPD}(b)). The transferred heat in the ferromagnetic layer will increase disorder in the magnetic moment, reducing its magnetization \cite{zhang_magnetization_1998} and therefore the magnetic field in the superconducting layer. This will reduce the vortex crossing barrier, leading to a detection event. 

In Fig. \ref{Fig:M_SNSPD}(c) we show the hybrid phase transition which occurs in the hybrid FM-SNSPD in response to an incident infrared photon. The hybrid FM-SNSPD detection mechanism is fundamentally different from a traditional SNSPD as it requires the magnetization of the ferromagnetic layer to transition to a more disordered state. Then, the vortex crossing event in the superconducting layer causes a transition to the normal state. By combining the effects of these two phase transitions, SNSPD noise can be reduced without significantly impacting sensitivity, enabling operation at higher temperatures. In Fig. \ref{Fig:M_SNSPD}(d) we use noise equivalent power (lower is more sensitive) to compare hybrid FM-SNSPD models to SNSPD experiments \cite{verma_single-photon_2021,chen_mid-infrared_2021,taylor_low-noise_2023,chang_efficient_2022,azem_mid-infrared_2024,marsili_efficient_2012} in the midwave infrared (MWIR) and longwave infrared (LWIR). Noise equivalent power (NEP) is given by \cite{hadfield_single-photon_2009}
\begin{equation}
    NEP = \frac{hf\sqrt{2DCR}}{SDE}
    \label{eq:NEP}
\end{equation}
where $h$ is Planck's constant, and $f$ is the photon frequency. We predict that hybrid FM-SNSPDs will have reduced noise, enabling higher temperature operation for infrared wavelengths.

\begin{figure*}[htb!]
\centering
\includegraphics[width=\textwidth]{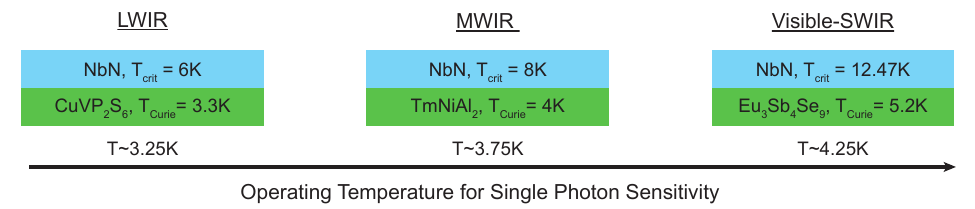}
\caption{Potential hybrid FM-SNSPD designs for LWIR, MWIR, visible-SWIR, and high temperature operations where $T$ is the operating temperature, $T_{crit}$ is the proximity effect reduced superconducting critical temperature and $T_{Curie}$ is the ferromagnet Curie temperature \cite{xu_low_2019,li_giant_2012,wang_magnetic_2020,jacob_enhanced_2021,hicks_giant_1969}.}
\label{Fig:Hybrid_Designs}
\vspace{-10pt}
\end{figure*}
Hybrid FM-SNSPDs requires a careful design procedure in which the temperature of operation and temperature dependent changes in the ferromagnet magnetization are simultaneously considered (see the Supplementary Material for further details \cite{supp}). Additionally, to enable improved sensitivity at longer wavelengths, the superconducting critical temperature may also need to be reduced \cite{colangelo_large-area_2022,verma_single-photon_2021,chen_mid-infrared_2021,taylor_low-noise_2023}, which can be achieved by increasing disorder in the superconducting layer \cite{zhang_saturating_2019}. In Fig. \ref{Fig:Hybrid_Designs} we present several potential designs for hybrid FM-SNSPDs including the ferromagnet material choices (green) based on their Curie temperature and the device operating temperature $T$. We note that the resulting designs have operating temperatures near the boiling point of liquid helium ($\sim4.2K$), with lower temperatures ($1-4.2K$) reachable through the addition of a helium compressor stage \cite{wang_closed-cycle_2014}. Additionally, a range of ferromagnetic materials are available with experimentally measured Curie temperatures between 1.5K and 100K (see Table \ref{tab1}). Several of these ferromagnets have also demonstrated perpendicular magnetic anisotropy (PMA), indicating the potential for magnetization along the out-of-plane axis. In the following sections we will explain the material properties necessary to achieve high operating temperatures in NbN-based hybrid FM-SNSPDs.

\begin{table*}[htb!]
\caption{List of potential ferromagnet candidates for hybrid FM-SNSPDs}\label{tab1}%
\resizebox{\textwidth}{!}{
\begin{tabular}{@{}l|ccccccccc@{}}
\botrule
  &LiHoF$_4$   & CuVP$_2$S$_6$  & TmNiAl$_2$ 
 & ErMn$_2$Si$_2$ 
 & Eu$_3$Sb$_4$Se$_9$ 
 & Eu$_2$SiO$_4$ 
 & ErNi$_2$ 
 & Ni$_x$Cu$_{1-x}$ 
 & Sb$_{2-x}$Cr$_x$Te$_3$ 
 \\
\hline
  $T_{Curie}$ (K)& 1.53 \cite{liu_ultralow-field_2023}  & 3.3 \cite{xie_cu-driven_2026} & 4.0 \cite{xu_low_2019} & 4.5 \cite{li_giant_2012} & 5.2 \cite{wang_magnetic_2020} & 5.8 \cite{mo_ferromagnetic_2025} & 6.5 \cite{cwik_structural_2018} & 9-100 \cite{jacob_enhanced_2021,hicks_giant_1969} & 15-37 \cite{dyck_low-temperature_2005}  \\

Anisotropy&  PMA  & - & - & - & PMA & - & - & PMA & PMA \\
\hline
\hline
\end{tabular}}
\end{table*}

\section{Vortex Engineering in Hybrid FM-SNSPD Devices}
\subsection{Time-Dependent Ginzburg Landau Analysis}
\begin{figure*}[hbt!]
\centering
\includegraphics[width=\textwidth]{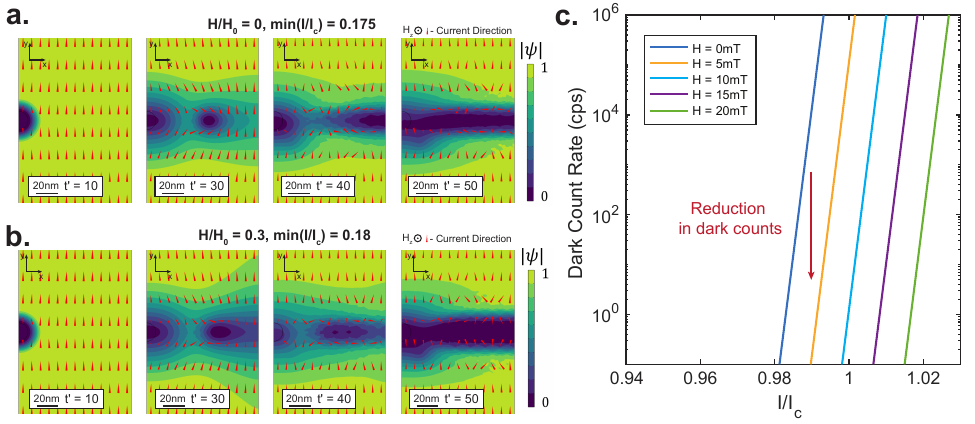}
\caption{Dark count behavior of SNSPDs and FM-SNSPDs. (a-b) Time dependent Ginzburg-Landau simulation of vortex crossing behavior for different bias currents and magnetic fields. Large positive magnetic fields are observed to require higher bias currents for single vortex crossing. (a) Vortex crossing-induced transition occurs at $I/I_{c} = 0.175$ for $H/H_0=0$ in an NbN SNSPD. (b) Vortex crossing-induced transition occurs at $I/I_{c} = 0.18$ for $H/H_0=0.3$ in an NbN SNSPD. (c) Dark count simulation from vortex crossing model demonstrating reduction in dark count rate as applied magnetic field increases. Note that relatively small increases in magnetic field can lead to significant reductions in dark count rate.}
\label{Fig:DarkCount}
\vspace{-10pt}
\end{figure*}

We begin by modeling the device behavior under a bias current and small applied magnetic field using time-dependent Ginzburg-Landau (TDGL) simulations (see the Supplementary Material for further details \cite{supp}) \cite{alstrom_magnetic_2011,PhysRevLett.83.2409,comsol}. We use these simulations to qualitatively analyze the vortex crossing behavior \cite{zotova_photon_2012}. In the TDGL simulation, a vortex is nucleated with a localized hotspot at the edge of the nanowire. Magnus forces from the applied current push against the vortex self-potential preventing movement across the width \cite{PhysRevB.83.144526,engel_detection_2015}. At some minimum current, the Magnus forces overcome the vortex self-potential leading to a vortex crossing event. In Fig. \ref{Fig:DarkCount}(a) we see that when no magnetic field is applied, a minimum current of $I/I_c = 0.175$ causes the vortex to cross the nanowire, leading to a phase transition. Here $I_c$ is the Ginzburg-Landau critical current under zero magnetic field, and $|\psi|$ is the magnitude of the superconducting order parameter. We find that for a magnetic field of $0.3H/H_0$ in the +z direction and current in the +y direction, the minimum current required to produce a phase transition is increased to $I/I_c = 0.18$ (see Fig. \ref{Fig:DarkCount}(b)). Using our vortex crossing model (based on Eq. \ref{eq:potential_function} and Eq. \ref{eq:DCR}) we find the dark count rate for varying magnetic fields (see Fig. \ref{Fig:DarkCount}(c)). From the model we see that a small increase in the magnetic field leads to a significant reduction in the dark count rate (DCR) for a fixed bias current.

\begin{figure*}[htb!]
\centering
\includegraphics[width=\textwidth]{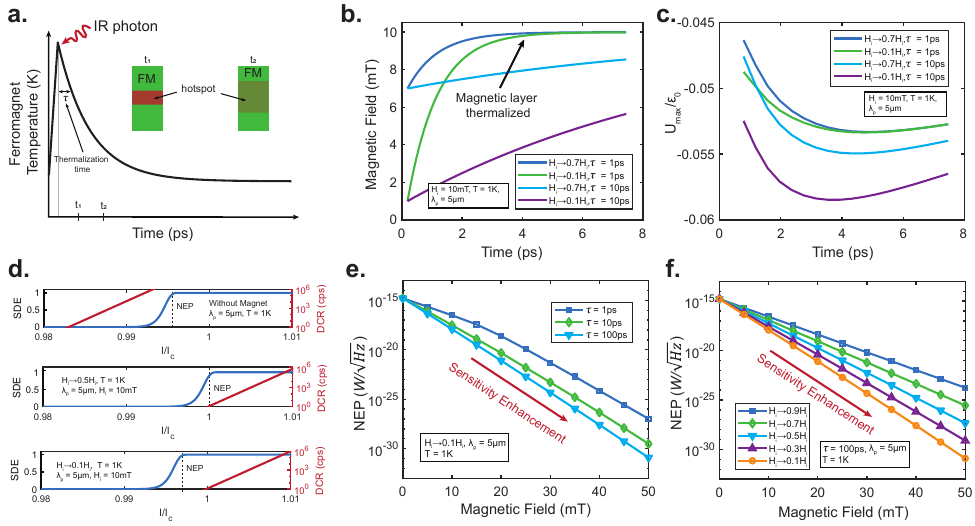}
\caption{Effect of time-varying ferromagnet temperature on SNSPD sensitivity. (a) An incident IR photon causes a change in the ferromagnet temperature which exponentially returns to the environment temperature with thermalization time $\tau$.  (b) Change in magnetic field for different reductions in magnetization and thermalization times. (c) Effect of different time-varying magnetic fields on the energy barrier. Note that larger thermal time constants lead to nearly constant changes in barrier. (d) Comparison of SDE and DCR for standard SNSPD (top) and FM-SNSPD with different reductions in magnetic field (middle, bottom). NEP is taken at the SDE plateau ($I/I_c$ where SDE = 0.99). The magnetic field moves both DCR and SDE plateau to larger $I/I_c$, while the reduction in magnetic field moves the SDE plateau back to it's original position. At larger magnetic fields, the SDE plateau occurs at bias current slightly larger than $I_c$. This limits the applicability of FM-SNSPDs to particular designs which have increased magnetic field reduction ($a$) for larger applied magnetic fields ($H_i$). (e) NEP versus magnetic field for different thermalization times demonstrating sensitivity enhancement. Similar trends are observed at larger magnetic fields. (f) NEP versus magnetic field for different reductions in magnetic field demonstrating sensitivity enhancement. }
\label{Fig:NEP}
\end{figure*}
\subsection{Effect of Photon-Induced Heating on Ferromagnet}
Next, we analyze the effect of temperature changes in the ferromagnet on SNSPD sensitivity. We model the change in ferromagnet temperature using a simple lumped capacitance model \cite{rogalski_infrared_2019}, taking the change in temperature to follow 
\begin{equation}
    T(t) = T_i +(T(0)-T_i)e^{-t/\tau}
    \label{Eq:temp}
\end{equation}
where $T(0)$ is the temperature after heating from the incident photon, $T_i$ is the temperature of the SNSPD prior to photon incidence, and $\tau$ is the thermal time constant \cite{lienhard_heat_2011,taneda_time-resolved_2007,besse_generation_2020}. The thermal time constant depends on the geometry and sound velocity of the ferromagnetic layer (see the Supplementary Material for further details \cite{supp}). After several $\tau$, the magnetic layer thermalizes to it's temperature prior to photon incidence (see Fig. \ref{Fig:NEP}(a)). We take the ferromagnet magnetization after heating to follow a similar trend as (Eq. \ref{Eq:temp}) \cite{taneda_time-resolved_2007}. Therefore, the time-varying magnetic field will follow
\begin{equation}
    H(t) = H_i(1-ae^{-t/\tau})
\end{equation}
where $H_i$ is the initial magnetic field, and $H_i(1-a)$ is the magnetic field after heating from the incident photon (see Fig. \ref{Fig:NEP}(b)). In Fig. \ref{Fig:NEP}(c) we show the effect of time-varying magnetic fields on the vortex crossing barrier $U_{max}$. The vortex crossing barrier typically changes in timescales of 5-15ps \cite{jahani_probabilistic_2020}, with the minimum barrier contributing the most to the count rate. Generally longer thermal time constants ($\tau \sim10-100ps$) act as a near constant change in the barrier in this 5-15ps window, while smaller thermal time constants ($\tau\sim1-10ps$) significantly modify the barrier until thermalization. 

\subsection{Effect of Hybrid Transition on Detector Sensitivity}
In Fig. \ref{Fig:NEP}(d) we compare the hybrid FM-SNSPD SDE (blue) and DCR (red) for various magnetic field profiles. We find that a small reduction in magnetic field can push the plateau region of the SDE close to the critical current (middle). As the reduction in magnetic field increases (bottom), the SDE displays similar plateau region as an SNSPD without a magnetic field (top). Meanwhile, the change in dark counts remain constant for devices with same initial magnetic fields (middle, bottom). We use the noise equivalent power metric (Eq. \ref{eq:NEP}) to compare different magnetic field conditions (i.e. different device designs). We take the NEP at the bias current of the initial SDE plateau (i.e. SDE$>$0.99, see Fig. \ref{Fig:NEP}(d)). Note that for large initial magnetic fields $(H_i)$ and small reductions in magnetic field $(a)$, the SDE plateau moves to bias currents slightly above $I_c$, limiting NEP. This limits the applicability of hybrid FM-SNSPDs to designs which have proportionally larger magnetic field reductions for larger initial magnetic fields.

In Fig. \ref{Fig:NEP}(e) we show the NEP versus magnetic field for different thermal time constants, and a large fixed change in magnetic field ($H\rightarrow0.1H$). The NEP reduces (i.e. sensitivity increases) as magnetic field increases for all thermal time constants and magnetic fields. Additionally, the dependence of NEP on magnetic field is approximately the same for all time constants. This indicates a range of time constants can be used to improve NEP. Next, in Fig. \ref{Fig:NEP}(f) we plot the NEP versus magnetic field for different reductions in magnetic field and a large fixed time constant ($\tau=100ps$). We find that in all cases sensitivity increases with magnetic field. 

\begin{figure*}[ht!]
\centering
\includegraphics[width=\textwidth]{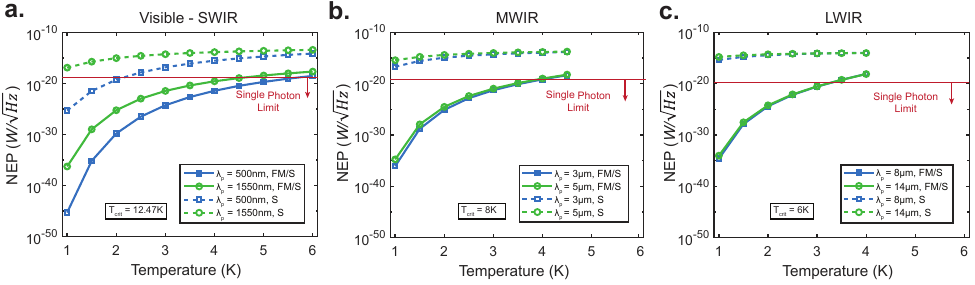}
\caption{Impact of ferromagnet/superconductor bilayers on infrared NEP. (a-c) Solid lines indicate hybrid FM-SNSPDs, while dashed lines indicate the same device designs without a magnetic layer. The hybrid FM-SNSPDs utilize a ferromagnet with $H_i = 40mT$, $H_i\rightarrow0.05H_i$, $\tau = 100ps$. NEP below the marked single photon limit indicates SNR $>$ 1 for incident photon flux of 1 photon/s at 1Hz readout bandwidth. (a) NEP versus temperature for visible to SWIR wavelengths demonstrating $\sim$ 4.25K operation for hybrid FM-SNSPDs. (b) NEP versus temperature for MWIR wavelengths demonstrating $\sim$ 3.75K operation for hybrid FM-SNSPDs. (c) NEP versus temperature for LWIR wavelengths demonstrating $\sim$ 3.25K operation for hybrid FM-SNSPDs. Significantly improved sensitivity/operating temperature is demonstrated in hybrid FM-SNSPD across all wavelengths.}
\label{Fig:NEP_wavelength}

\end{figure*}

\subsection{Effect of Hybrid Transition on Operating Temperature}
Next, we look at the effect of the FM-SNSPD design on operating temperature. In Fig. \ref{Fig:NEP_wavelength} we plot the temperature dependent NEP of hybrid FM-SNSPDs with particular design properties ($H_i = 40mT, H\rightarrow.05H, \tau = 100ps$). We additionally utilize reduced superconducting critical temperatures as mentioned in Fig. \ref{Fig:Hybrid_Designs} to enable sensitive operation at longer wavelengths. Note that in all bands, significant improvements in NEP are achieved for the hybrid FM-SNSPD case (solid lines) compared to the standard SNSPD (dashed lines).  In red we mark the single photon limit which we define as the power at which signal to noise ratio (SNR) $=$ 1 for incident photon flux of 1 photon/s at 1Hz readout bandwidth. We find that single photon operating temperatures of 4.25K, 3.75K, and 3.25K are possible for 1550nm, 5$\mu$m, and 14$\mu$m light respectively (see Fig. \ref{Fig:NEP_wavelength}a, b, and c). Note that the visible-SWIR bandwidth can be operated with open cycle liquid helium cryostats (4.2K). Meanwhile the MWIR-LWIR bandwidth can be operated by a closed cycle liquid helium cryostat with a helium compressor (1-4.2K) \cite{wang_closed-cycle_2014}.

\begin{figure*}[ht!]
\centering
\includegraphics[width=\textwidth]{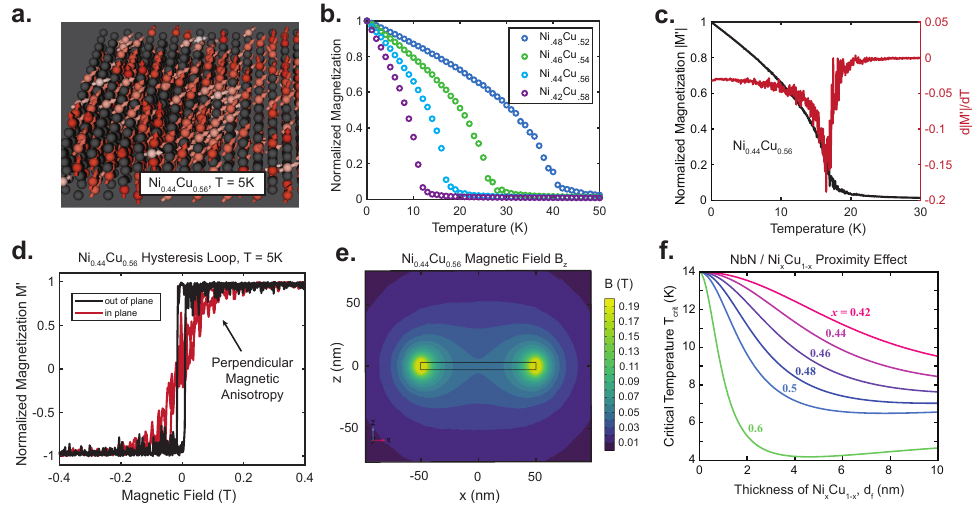}
\caption{Synthetic ferromagnet design. (a) Atomistic spin dynamics (ASD) simulation of Ni$_{0.44}$Cu$_{0.56}$ at $T=5K$. Gray atoms are non-magnetic Cu, while red$\rightarrow$white atoms are Ni with out-of-plane$\rightarrow$in-plane orientations. (b) ASD simulation of magnetization versus temperature for different nickel and copper concentrations. The Curie temperature is observed to decrease as copper concentration increases. (c) Normalized magnetization curve (black) and derivative of magnetization versus temperature (red) for Ni$_{0.44}$Cu$_{0.56}$, demonstrating optimal temperature biasing point. (d) Hysteresis loop of Ni$_{0.44}$Cu$_{0.56}$ at $T=5K$ obtained from ASD simulations, demonstrating perpendicular magnetic anisotropy. (e) Out-of-plane magnetic field profile of 6nm Ni$_{0.44}$Cu$_{0.56}$ layer at $T=5K$. (f) Effect of various NiCu alloys on superconducting critical temperature. The proximity effect plays a minor role on the critical temperature of NbN for thin magnetic layers with large concentrations of copper. These synthetic ferromagnets are capable of producing significant magnetic fields in the superconducting layer, providing a method for ferromagnet-based vortex engineering.}
\label{Fig:Ferromagnet}
\vspace{-10pt}
\end{figure*}

\section{Designing a Ferromagnetic Layer for FM-SNSPDs}
As we've shown in the previous section, hybrid FM-SNSPDs have the potential to increase operating temperatures of MWIR/LWIR single photon detectors. However, the designs capable of achieving improved sensitivity require the ferromagnetic layer to have particular characteristics. Specifically, we require the ferromagnet to produce relatively small magnetic fields (10-50mT) in the superconducting layer, have a perpendicular magnetic anisotropy, and have a Curie temperature close to the operating temperature. These characteristics can be modeled using a combination of atomistic spin dynamics simulations and finite element simulations \cite{evans2014atomistic,vampire,comsol}. Additionally, we need to determine the effect of the ferromagnetic layer on the superconducting critical temperature, which can be obtained from the Usadel equations. In this paper we focus our modeling efforts on the ferromagnetic alloy $Ni_xCu_{1-x}$, although a similar approach could be used to model the behavior of other ferromagnetic materials.

\subsection{Simulating Magnetic Properties of Ni$_x$Cu$_{1-x}$}
In our atomistic spin dynamics (ASD) simulations we describe the spin evolution using a Landau-Lifshitz-Gilbert equation with a thermal field term governed by Langevin dynamics (see the Supplementary Material for further details \cite{supp})\cite{evans2014atomistic,lyberatos1993method,nowak2005spin,1060329,garcia1998langevin,vampire,colarieti2003origin,ikeda1984origin,ahmad1974electrical}. We use ASD to simulate the random alloy of NiCu with Ni as the alloy host. The thickness is set to 6 nm while the x and y boundaries are set to periodic. The red/white atoms shown in Fig. \ref{Fig:Ferromagnet}(a) represent Ni and the gray atoms represent Cu. During the simulation, Cu is treated as a nonmagnetic atom randomly distributed in the system. These nonmagnetic impurities break down the exchange coupling of the magnetic Ni atoms \cite{hicks_giant_1969,aitken_spin-glass-like_1979,ododo1977critical}, resulting in a significant decrease in the Curie temperature due to the suppression of magnetic properties. Fig. \ref{Fig:Ferromagnet}(b) shows the effect of changing Cu concentration on the Curie temperature $T_{Curie}$ of the ferromagnetic alloy.  At 60\% Cu the alloy  transitions from a ferromagnet phase to a spin-glass phase \cite{aitken_spin-glass-like_1979}.

From our ASD simulation we find that Ni$_{x}$Cu$_{1-x}$ alloys can achieve a tunable Curie temperature of at least 10-40K (See Fig. \ref{Fig:Ferromagnet}(b)), matching closely with experimental results \cite{hicks_giant_1969}. In particular, Ni$_{0.44}$Cu$_{0.56}$ is a potentially favorable concentration for FM-SNSPDs with a  $T_{Curie}$ of 17K. In Fig. \ref{Fig:Ferromagnet}(c) we plot the derivative of Ni$_{0.44}$Cu$_{0.56}$ magnetization versus temperature. The maximum derivative represents the biasing temperature with the largest reduction in magnetic field. However, significant reductions in magnetization are possible even at lower operating temperatures, such as those used in NbN-based hybrid FM-SNSPDs. Additionally, operation at lower temperatures may be required to maintain initial magnetic field strength ($H_i$) in the superconducting layer. In Fig. \ref{Fig:Ferromagnet}(d) we use ASD to simulate the hysteresis of Ni$_{0.44}$Cu$_{0.56}$ for in-plane (x,y) and out-of-plane (z) applied magnetic field directions. We find that Ni$_{0.44}$Cu$_{0.56}$ exhibits perpendicular magnetic anisotropy based on the small coercivity in the out-of-plane direction. Next, we model the magnetic field based on the magnetization obtained from ASD simulations (see Fig. \ref{Fig:Ferromagnet}(e)), taking the ferromagnetic layer to be 100nm wide and 6nm thick. Large field gradients occur at the edges, but a superconductor placed at the center with half the width of the ferromagnet would see a more uniform field of $\sim$40mT. 

\subsection{Proximity Effect of Ni$_x$Cu$_{1-x}$}
It is well known that integrating a superconducting material with a non-superconducting material can influence the superconducting critical temperature. To address this, we consider the influence of the ferromagnetic layer on the superconducting critical temperature due to the proximity effect. The ferromagnet/superconductor bilayer's critical temperature $T_{\rm CS}$ can be found by solving the self-consistency equation obtained from solving the Usadel equations \cite{usadel1970generalized, buzdin2005proximity}. We use material parameters from literature \cite{charaev_magnetic-field_2019,zdravkov_reentrant_2006,ryazanov_proximity_2003} and the exchange potential $E_{xc}$ is obtained from fits to experimental Curie temperatures \cite{jacob_enhanced_2021,ododo1977critical,evans2014atomistic}. The exchange potential can also be calculated from ab-initio simulations \cite{he_tb2j_2021}. The self-consistency equation can be written in the following form:
\begin{equation}
    \ln\frac{T_{\rm CS}}{T_c} = {\rm Re}\,\psi\left(\frac{1}{2}+\frac{\gamma}{2}\frac{\xi_s}{d_s}\frac{1}{\gamma_b+B_f}\frac{T_{\rm CS}}{T_C}\right) - \psi\left(\frac{1}{2}\right)
\end{equation}
where,
\begin{subequations}
\begin{align}
    B_f &= \left[k_f \xi_f \tanh\left(k_f d_f\right)\right]^{-1};  \\
    k_f &= \frac{1}{\xi_f} \sqrt{\frac{E_{xc}}{\pi T_{\rm CS}}}; \\
    \gamma_b &= \frac{R_b\mathcal{A}}{\rho_f\xi_f}; \\
    \gamma &= \frac{\rho_s}{\rho_f}\frac{\xi_s}{\xi_f},
\end{align}
\end{subequations}
$\psi$ is the digamma function, $R_b$ is the resistance of the ferromagnet/superconductor interface, $\mathcal{A}$ is the area of the interface, $d_f$ is the thickness of the ferromagnetic layer, $d_s$ is the thickness of the superconducting layer, $\xi_f$ is the ferromagnet coherence length, $\xi_s$ is the superconducting coherence length, $\rho_f$ is the ferromagnet normal state resistivity, $\rho_s$ is the superconductor normal state resistivity, and $T_c$ is the superconductor critical temperature. In Fig. \ref{Fig:Ferromagnet}(f), we plot $T_{\rm CS}$ as a function of the ferromagnetic layer thickness and copper concentration for an NbN/Ni$_{x}$Cu$_{1-x}$ interface (see the Supplementary Material for further details \cite{supp}). Note that $T_{\rm CS}$ displays a nonmonotonic behavior as the thickness of the ferromagnetic layer increases. For large concentrations of copper we find that the change in critical temperature is fairly minimal, even for ferromagnetic layers of 3-6nm in thickness.

\section{Conclusions}
In conclusion, we have demonstrated a vortex-engineering approach to SNSPD design based on hybrid ferromagnet-superconductor phase transitions. We find that ferromagnets with particular temperature dependent characteristics can reduce dark noise without significantly affecting detection efficiency.  Based on this approach, we have presented several potential designs for MWIR/LWIR single photon detectors utilizing low Curie temperature ferromagnets and reduced energy gap superconductors. These hybrid FM-SNSPD devices can enable larger operating temperatures of 3.25-3.75K, leading to significantly simpler cooling systems for infrared single photon detection.

\section*{DATA AVAILABILITY}
The data that support the findings of this article are not publicly available. The data are available upon reasonable request from the authors.

\begin{acknowledgments}
This work was funded by the DARPA SynQuaNon program.
\end{acknowledgments}

\section*{Appendix}
Detailed description of simulation methods for time-varying vortex crossing barrier, time-dependent Ginzburg-Landau theory, atomistic spin dynamics simulations, and proximity effect calculations. Analysis of single-photon induced changes in magnetization and analysis of thermalization timescales in ferromagnet/superconductor bilayer.
\clearpage

\appendix

\onecolumngrid
\section{Time-Varying Vortex Crossing Barrier}
We simulate the time-varying vortex crossing barrier using a quasiparticle diffusion process \cite{engel_detection_2015,jahani_probabilistic_2020} (Eq. \ref{eq:quasi_1} and \ref{eq:quasi_2}) modeled via finite-difference methods. Using the quasiparticle distribution densities found from the diffusion model we can calculate the time-varying current distribution from Eq. \ref{eq:current} \cite{engel_detection_2015,jahani_probabilistic_2020}, which impacts the vortex crossing barrier (Eq. \ref{eq:VortexBarDist}) \cite{engel_detection_2015,jahani_probabilistic_2020}. The hot electron and quasiparticle diffusion are found from the following equations \cite{engel_detection_2015,jahani_probabilistic_2020}
\begin{equation}
     \frac{\partial C_e(\mathbf{r}, t)}{\partial t} = D_e\nabla^2C_e(\mathbf{r}, t) 
     \label{eq:quasi_1}
 \end{equation}
 \begin{equation}
     \frac{\partial C_{qp}(\mathbf{r}, t)}{\partial t} = D_{qp}\nabla^2C_{qp}(\mathbf{r}, t) - \frac{C_{qp}(\mathbf{r}, t)}{\tau_r} + \frac{\varsigma h\nu}{\Delta\tau_{qp}}\left(\frac{n_{se,0}-C_{qp}(\mathbf{r}, t)}{n_{se,0}}\right)e^{-t/\tau_{qp}}C_e(\mathbf{r}, t)
     \label{eq:quasi_2}
 \end{equation}
where $C_e(\mathbf{r}, t)$, $C_{qp}(\mathbf{r}, t)$ are the hot electron and quasi-particle distribution densities respectively. $D_e$ is the the hot-electron diffusion coefficient, $D_{qp}$ is the quasi-particle diffusion coefficient, $\tau_r$ is the recombination time, $\tau_{qp}$ is the quasiparticle lifetime, $\varsigma=0.25$ is the quasiparticle multiplication efficiency, $h\nu$ is the incident photon energy, $\Delta$ is the superconducting bandgap, and $n_{se,0}$ is the density of superconducting electrons prior to photon absorption. The current distribution can be calculated using the quasiparticle distribution density from the following equations \cite{engel_detection_2015,jahani_probabilistic_2020}
\begin{equation}
    \nabla\cdot\mathbf{j}(\mathbf{r}, t) = \nabla\cdot\left(\frac{\hbar}{m}n_{se}(\mathbf{r}, t)\nabla\varphi(\mathbf{r}, t)\right) = 0
    \label{eq:current}
\end{equation}
 \begin{equation}
    n_{se}=n_{se,0}-C_{qp}
\end{equation}
where $\mathbf{j}(\mathbf{r}, t)$ is the current distribution, $n_{se}(\mathbf{r}, t)$ is the density of superconducting electrons after photon absorption, $\varphi$ is the phase of the superconducting order parameter, $\hbar$ is the reduced Planck's constant, and $m$ is the mass of an electron. The vortex crossing barrier is then given by \cite{engel_detection_2015,jahani_probabilistic_2020}
\begin{eqnarray}
\label{eq:VortexBarDist}
\frac{U_{max}(t)}{\epsilon_0}= \nonumber&&\max_{x_\nu}\left[\frac{\pi}{W}\int^{x_v}_{\frac{\xi-W}{2}}
\frac{n_{se}(x',t)}{n_{se,0}}\tan\left(\frac{\pi x'}{W}\right)dx'\right.\\
&&-\frac{2W}{I_c\exp(1)\xi}\int^{x_v}_{\frac{-W}{2}}
\left.\frac{n_{se}(x',t)}{n_{se,0}}j_y(x',t)dx'\right]\\\nonumber
\end{eqnarray}
 where $W$ is the nanowire width, $d$ is the  nanowire thickness, $x_v$ is the vortex position, and $\xi$ is the coherence length. The characteristic vortex energy $\epsilon_0$ is given by Eq. 2 in the main text. We utilize the material parameters in Table \ref{tab2} (taken from \cite{engel_detection_2015,charaev_magnetic-field_2019}) to model the barriers found in the main text. The superconducting bandgap, magnetic penetration depth, coherence length, quasiparticle diffusion constant, and vortex attempt rate are found from the following temperature dependent equations \cite{tinkham_introduction_2015,engel_numerical_2013,he_unified_2025}:
\begin{equation}
    \Delta_0 = c_1k_BT_c
\end{equation}
\begin{equation}
    \Delta(T) = \Delta_0\tanh\left(1.74\sqrt{T_c/T-1}\right)
\end{equation}
\begin{equation}
    \lambda^{-2}(T) = \lambda^{-2}_0(0)\frac{\Delta(T)}{\Delta_0}\tanh(\Delta(T)/2k_BT)
\end{equation}

\begin{equation}
    \xi(T) = \sqrt{\Phi_0/(2\pi H_{c2}(T))}
\end{equation}
\begin{equation}
    D_{qp} = 0.6D_e\sqrt{T/T_c}
\end{equation}
\begin{equation}
\alpha_v(T)=\alpha_0\exp(-\beta_1/k_BT)
\end{equation}
\begin{table}[h]
\centering
\begin{tabular}{l | l}
Parameters & Values \\
\hline \hline
Nanowire Width $W$ & 100 nm \\
Superconductor Thickness $d$ & 10 nm \\
Superconducting Bandgap $\Delta_0$ & 1.03-2.15 meV \\
Superconducting Bandgap constant $c_1$ & 2 \\
Coherence Length $\xi_0$ & 4.1 nm \\
Magnetic Penetration Depth $\lambda_0$ & 289 nm \\
Electron Diffusion Constant $D_e$ & 0.54 $cm^{2}/s$ \\
Quasiparticle Lifetime $\tau_{qp}$ & 1.6 ps \\
Recombination Time $\tau_{r}$ & 1 ns \\
$\alpha_0$ &  574 kHz \\
$\beta_1$ & 0.175 meV \\
\end{tabular}
\caption{Vortex Crossing Simulation Parameters}
\label{tab2}
\end{table}

\section{Photon-Induced Magnetization Change}
A key aspect of the ferromagnet design is the reduction in magnetization due to a single incident photon. In order to estimate the change in temperature induced by a single photon, we first find the estimated change in temperature induced in the superconductor based on the quasiparticle diffusion model \cite{engel_numerical_2013,he_unified_2025}. In the quasiparticle diffusion model, the density of superconducting electrons is reduced from it's equilibrium value due to the hot-spot induced quasiparticle multiplication process. This reduced density can be related to the superconducting energy gap using Ginzburg-Landau theory
\begin{equation}
    n_{se}(T)\propto|\psi(T)|^2\propto\Delta(T)^2 
\end{equation}

\begin{equation}
    \sqrt{\frac{n_{se}(T+\Delta T)}{n_{se}(T)}} =\frac{|\psi(T+\Delta T)|}{|\psi(T)|}=\frac{\Delta(T+\Delta T)}{\Delta(T)}
\end{equation}
where $n_{se}$ is the density of superconducting electrons, $\psi$ is the order parameter, $\Delta$ is the superconducting energy gap, $T$ is temperature, and $\Delta T$ is the maximum change in temperature due to the incident photon \cite{tinkham_introduction_2015,engel_numerical_2013,he_unified_2025}. The temperature dependence of the superconducting energy gap is given by
\begin{equation}
    \Delta(T)=\Delta(0)\tanh\left(1.74\sqrt{T_c/T}-1\right)
\end{equation}
where $T_c$ is the superconducting critical temperature \cite{tinkham_introduction_2015}. Thus, a change in the superconducting electron density can be related to a change in the superconducting energy gap, which gives us a change in temperature $\Delta T$. We plot this change in temperature versus wavelength in Fig. \ref{fig: Temp_change}a with the same device parameters as specified in Supplementary Section A. Note that the changes in temperature follow a $1/\lambda$ dependence, as expected by the reduction in photon energy. Based on this temperature change, we can then calculate the change in magnetization, which is given by Bloch's law
\begin{equation}
    M(T)=M(0)\left(1-\left(\frac{T}{T_{Curie}}\right)^{3/2}\right)
\end{equation}
where $M$ is the magnetization, and $T_{Curie}$ is the Curie temperature \cite{zhang_magnetization_1998}. Thus the normalized reduction in magnetization is given by
\begin{equation}
    1-a=\frac{M(T+\Delta T)}{M(T)}.
\end{equation}
In Fig. \ref{fig: Temp_change}b we plot the normalized reduction in magnetization versus Curie temperature for the FM-SNSPD designs specified in Fig. 2 of the main text. For the field reduction of $H_i\rightarrow0.05H_i$, as specified in Fig. 5 of the main text, a Curie temperature below 5.493K, 4.006K, and 3.321K are required for the VIS-SWIR, MWIR, and LWIR deisgns respectively. Therefore the Curie temperatures of the ferromagnets specified in Fig. 2 of the main text meet the requirements of an FM-SNSPD.
\begin{figure}[ht!]
\centering
\includegraphics[width=0.8\textwidth]{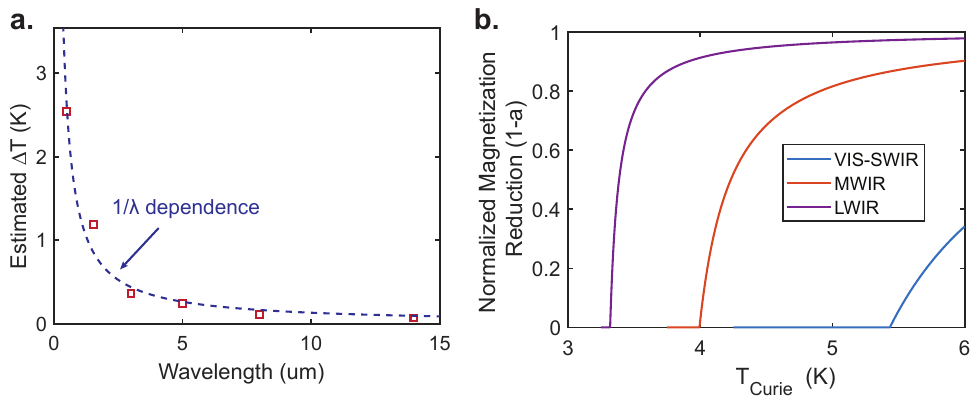}
\caption{Estimated effect of a single photon on the change in magnetization. (a) The red squares indicate estimated change in temperature due to a single photon as a function of wavelength. The change in temperature is demonstrated to show a 1/$\lambda$ dependence as expected by the decreasing photon energy. (b) Normalized magnetization reduction $(1-a)$ as a function of ferromagnet Curie temperature for the SNSPD designs specified in Fig. 2 of the main text.}
\label{fig: Temp_change}

\end{figure}

\section{Time-Dependent Ginzburg Landau Simulations}
The GL free energy in its dimensionless form is given by
\begin{equation}
     F_{GL}=\int\left[\left|\psi^*(\vec{\nabla}/i-\vec{A})\psi\right|^2-|\psi|^2+\frac{\kappa^2}{2}|\psi|^4+\left|\vec{\nabla}\times\vec{A}\right|^2\right]d^3x.
     \label{eq:GLfree}
 \end{equation}
The normalized TDGL equations for conventional superconductors (i.e. NbN, WSi, etc.) are given by
  \begin{equation}
     \left(\frac{\partial}{\partial t}+i\kappa\varphi\right)\psi=-\left(\frac{i}{\kappa}\vec{\nabla}+\vec{A}\right)^2\psi+\left(1-\frac{T}{T_c}\right)\psi-\left|\psi\right|^2\psi
     \label{eq:order_parameter_fin}
 \end{equation}
 \begin{equation}
     \sigma\left(\frac{\partial\vec{A}}{\partial t}+\vec{\nabla}\varphi\right)=\frac{1}{2i\kappa}\left(\psi^\ast\vec{\nabla}\psi-\psi\vec{\nabla}\psi^\ast\right)-\left|\psi\right|^2\vec{A}-\vec{J_b}-\vec{\nabla}\times(\vec{\nabla}\times\vec{A}-\vec{B_a})
     \label{eq:A_eqn_fin}
 \end{equation}
where $\psi$ is the superconducting order parameter, $\kappa=\lambda/\xi$ denotes the ratio of penetration depth $\lambda$ to coherence length $\xi$, $\vec{A}$ is the magnetic vector potential, $\varphi$ is the electric potential, $T$ is the operating temperature, $T_c$ is the superconducting critical temperature, $\sigma$ is the conductivity of the normal current, $\vec{J_b}$ is the bias current density, and $\vec{B_a}$ is the applied magnetic field. Additional boundary conditions for current flow have been applied in our approach such that $\psi^*\vec{\nabla}\psi\cdot\vec{n}=i\vec{J_b}$ at the boundaries where current flows, and $\vec{\nabla}\psi\cdot\vec{n}=0$ at the other boundaries. The order parameter $\psi$ is calculated using the TDGL equations by transforming Eq. \ref{eq:order_parameter_fin} and \ref{eq:A_eqn_fin} into coupled partial differential equations (PDEs). Those PDEs are solved through finite element methods by modeling in the COMSOL Multiphysics\textsuperscript{\textregistered} software \cite{comsol}.

\section{Thermalization Timescales}
A hot spot generated in the superconducting layer will diffuse to the ferromagnetic layer in a characteristic time scale $t\approx L^2/4D$ where $D$ is the electron diffusion constant  \cite{lienhard_heat_2011}. For a 10nm superconducting layer and a diffusion constant of 54$nm^2/ps$\cite{charaev_magnetic-field_2019} we can expect the diffusion time to be approximately 0.46 ps. For NbN the delay will remain less than 2ps for superconductor thicknesses less than 20nm. 

\begin{figure}[ht!]
\centering
\includegraphics[width=0.8\textwidth]{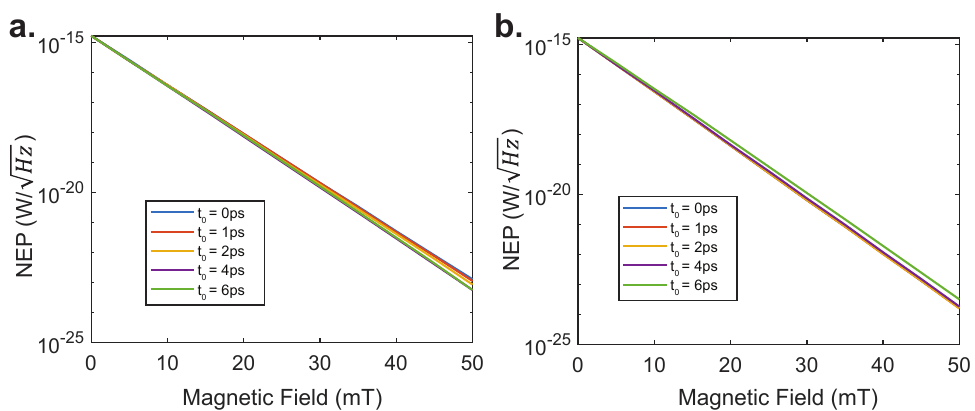}
\caption{Effect of delay on sensitivity. (a) NEP versus magnetic field for different delays and $\tau = 1ps$. (b) NEP versus magnetic field for different delays and $\tau = 100ps$.}
\label{fig: Delay_comparison}

\end{figure}

The delayed time-varying magnetic field can be modeled as $H(t)=H_i(1-au(t-t_0)e^{-(t-t_0)/\tau})$ where $t_0$ is the delay and $u(t)$ is a unit step function. We model the effects of the delay on NEP in Fig. \ref{fig: Delay_comparison}.  Note that small delays slightly enhance sensitivity for magnetic layers with small thermalization times $\tau$ (see Fig. \ref{fig: Delay_comparison}a, $\tau$ = 1ps), but have little effect on field profiles with long $\tau$ (see Fig. \ref{fig: Delay_comparison}b, $\tau$ = 100ps). Delays longer than 4ps reduce sensitivity as the changes in barrier from the superconductor hotspot become mismatched with the changes in field. Overall, delay plays a minor role on the NEP. To simplify the assumptions in our model we assume the delay is zero in the main text.

The ferromagnet thermalization time $\tau$ is primarily determined by electron-phonon coupling and by the phonon escape time \cite{taneda_time-resolved_2007}. Therefore, although we take a simple exponential model, experiments suggest that two or three exponential timescales may be present in ferromagnet/superconducting bilayers \cite{taneda_time-resolved_2007}. We will assume the thermalization time is primarily limited by the phonon escape time, as has been demonstrated in bilayer experiments. The phonon escape time can be calculated using
\begin{equation}
    \tau_{es} = 4d/\eta v_s
\end{equation}
where $d$ is the ferromagnet thickness, $\eta$ is the acoustic transmission probability and $v_s$ is the sound velocity. Taking $d = 6nm$, $\eta = 0.25$ \cite{taneda_time-resolved_2007}, and $v_s = 5.6\times10^3m/s$ \cite{besse_generation_2020}, we have $\tau_{es}=17ps$ for a Ni ferromagnet. This is on the same scale as the thermalization times used in our simulations.

\section{Atomistic Spin Dynamics}
The simulations of ferromagnetic NiCu are performed using \ssmall VAMPIRE \normalsize \cite{vampire} which utilizes an atomistic spin dynamics (ASD) model \cite{evans2014atomistic}. The basis of the model is a classical spin Hamiltonian, which describes the fundamental spin-dependent interaction energy at the atomic level as  \cite{evans2014atomistic}
\begin{equation}
    \mathcal{H}=-\sum_{i<j}{J_{ij}\mathbf{S}_i\cdot\mathbf{S}_j-k_u\sum_{i}{{{(\mathbf{S}}_i\cdot\mathbf{e}_i)}^2\ -\mu_S\sum_{i}{\mathbf{H}_{app}\cdot\mathbf{S}_i}}}.
\end{equation}
Here the first term describes the Heisenberg exchange, where \(J_{ij}\) is the exchange energy, \(\mathbf{S}_i\) is a unit vector that describes the orientation of the local spin moment, and \(\mathbf{S}_j\) is the spin moment direction of neighboring atoms. The second term describes the uniaxial anisotropy contribution, where \(k_u\) is the uniaxial anisotropy energy and \(\mathbf{e}_i\) is a unit vector along the magnetic easy axis. The last term accounts for coupling of the spin system to an externally applied field \(\mathbf{H}_{app}\) depending on its atomic magnetic moment \(\mu_S\).

\ssmall VAMPIRE \normalsize uses the Landau-Lifshitz-Gilbert (LLG) equation to describe the dynamics of atomic spins, which is solved using the Monte-Carlo method. In general, the LLG equation is expressed as \cite{evans2014atomistic}
\begin{equation}
    \frac{\partial\mathbf{S}_i}{\partial t}=-\frac{\gamma}{\left(1+\lambda^2\right)}\left[\mathbf{S}_i\times\mathbf{H}_{eff}^i+\lambda\mathbf{S}_i\times\left(\mathbf{S}_i\times\mathbf{H}_{eff}^i\right)\right]
\end{equation}
where \(\gamma\) is the gyromagnetic ratio, \(\lambda\) is the Gilbert damping parameter, and \(H_{eff}^i\) is the net magnetic field on each spin. In its standard form, the atomistic LLG equation describes the interaction of an atomic spin moment \(i\) with an effective magnetic field at zero temperature. The net field for this case is expressed in terms of the complete spin Hamiltonian as  \cite{evans2014atomistic}
\begin{equation}
    \mathbf{H}_{eff}^i=-\frac{1}{\mu_S}\frac{\partial\mathcal{H}}{\partial\mathbf{S}_i}.
\end{equation}

At a finite temperature, the presence of thermal effects leads to thermodynamic fluctuations in the spin moments. If these fluctuations surpass the influence of the exchange interaction, the magnetic material will undergo a ferromagnetic-paramagnetic phase transition \cite{evans2014atomistic}. We utilize Langevin dynamics to account for these finite temperature effects  \cite{1060329} which is widely employed in spin dynamics simulations \cite{garcia1998langevin,lyberatos1993method,nowak2005spin}. Thermal fluctuations at each atomic site are represented by a Gaussian white noise term in the effective field. As the temperature rises, the increasing magnitude and width of the Gaussian distribution leads to stronger thermal fluctuations. The effective thermal field \(H_{th}^i\)  from Langevin dynamics is given by \cite{evans2014atomistic}
\begin{equation}
    \mathbf{H}_{th}^i=\ \mathbf{\Gamma}(t)\sqrt{\frac{2\lambda k_BT}{\gamma\mu_S\Delta t}}
\end{equation}
where \(k_B\) is the Boltzmann constant, and \(\Delta t\) is the integration time step. \(\mathbf{\Gamma}(t)\) is the Gaussian distribution in three dimensions with a mean of zero, and standard deviation given by \cite{garcia1998langevin}
\begin{equation}
    \sigma = \sqrt{2\frac{\lambda}{1+\lambda^2}\frac{k_BT}{\gamma \mu_s}\Delta t}.
\end{equation}
The effective field for a finite temperature is given by \cite{evans2014atomistic}
\begin{equation}
    \mathbf{H}_{eff}^i=-\frac{1}{\mu_S}\frac{\partial\mathcal{H}}{\partial\mathbf{S}_i}+\mathbf{H}_{th}^i.
\end{equation}

\begin{table}[h]
\centering
\begin{tabular}{l | l}
Parameters & Values \\
\hline \hline
Unit cell size & 3.524 Å \\
Exchange energy \(J_{ij}\) (Ni) & 2.757 E-21 J/link \\
Exchange energy \(J_{ij}\) (\(\mathrm{Ni_{0.44}Cu_{0.56}}\)) & 0.2319 E-21 J/link \\
Atomic spin moment \(\mu_S\) (Ni) & 0.606 \(\mu_B\)/atom \\
Atomic spin moment \(\mu_S\) (\(\mathrm{Ni_{0.44}Cu_{0.56}}\)) & 0.0510 \(\mu_B\)/atom \\
Anisotropy energy  \(k_u\) & 5.47 E-26 J/atom \\
Simulation integrator & Monte-Carlo \\
\end{tabular}
\caption{ASD Simulation Parameters}
\label{tab:ASD_parameters}
\end{table}

We use the parameters in Table \ref{tab:ASD_parameters} to find the magnetic properties of NiCu alloy for different concentrations. These parameters are either taken from existing literature data \cite{evans2014atomistic,ododo1977critical,colarieti2003origin}
or derived from mean-field approaches and density functional theory calculations. From our ASD simulation we find that ferromagnetic NiCu alloy can reach a tunable Curie temperature $<$20K with Ni$<$44\% (Cu$>$56\%). \(\mathrm{Ni_{0.44}Cu_{0.56}}\) is a favorable concentration for hybrid FM-SNSPD designs with a $T_{Curie}$ of 17K. The effects of changing the alloy composition on  the Curie temperature are shown in Figure \ref{fig: ASD_comparison} below, matching closely with experiments \cite{ododo1977critical,jacob_enhanced_2021,ikeda1984origin,ahmad1974electrical}. 

\begin{figure}[ht!]
\centering
\includegraphics[width=0.8\textwidth]{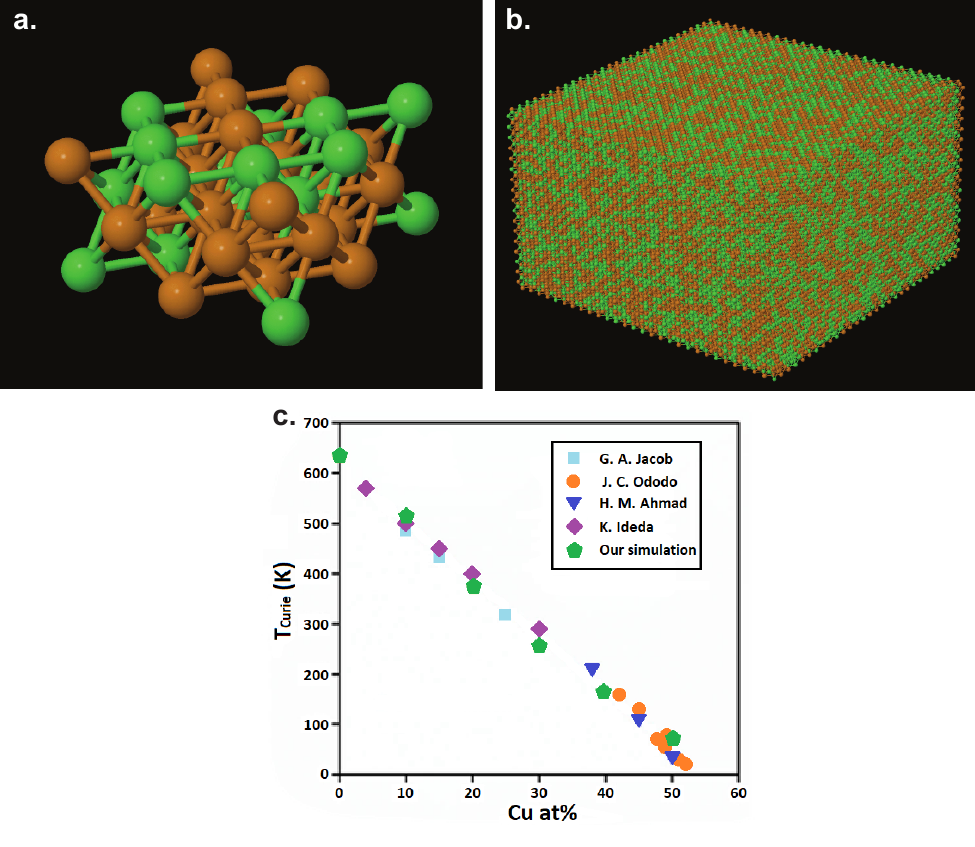}
\caption{Atomistic spin dynamics simulations. (a) FCC lattice structure with Ni (green) and Cu (orange) atoms. (b) System schematic with Cu atoms randomly distributed in the Ni alloy. (c) $T_{Curie}$ with various concentrations of Cu, results are either theoretical or experimental from literature \cite{ododo1977critical,jacob_enhanced_2021,ikeda1984origin,ahmad1974electrical}.}
\label{fig: ASD_comparison}

\end{figure}

\section{Proximity Effect Calculation}
\begin{figure}
    \centering
    \includegraphics[width=0.5\linewidth]{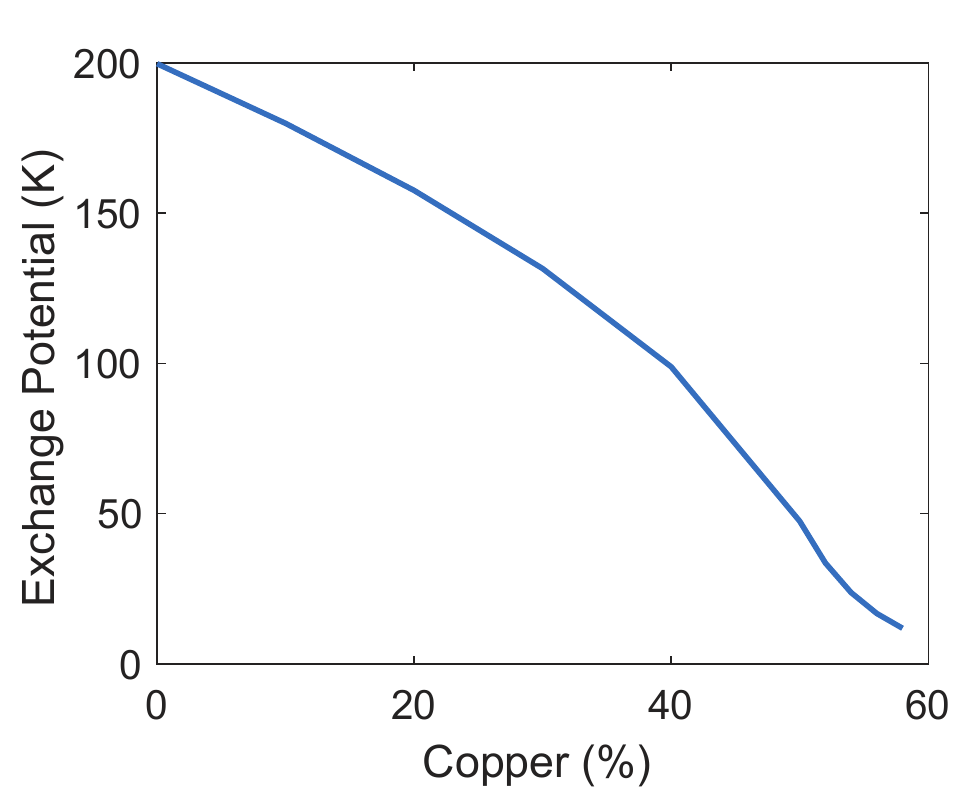}
    \caption{Exchange potential of Ni$_{x}$Cu$_{1-x}$ is plotted as a function of the copper concentration $x$. The exchange potential is obtained from fits to experimental Curie temperatures \cite{evans2014atomistic,jacob_enhanced_2021,ododo1977critical}.}
    \label{fig:exchange_potential}
\end{figure}

The proximity effect is dependent on the exchange energy $J_{xc}$ which we obtained from fits to concentration dependent Curie temperatures of Ni$_{x}$Cu$_{1-x}$ in literature \cite{evans2014atomistic,jacob_enhanced_2021,ododo1977critical}. In Fig. \ref{fig:exchange_potential} we plot the exchange potential $E_{xc}=J_{xc}/k_B$ versus copper concentration for the ferromagnetic alloy Ni$_{x}$Cu$_{1-x}$. We calculate the proximity effect for different concentrations using this exchange potential and the parameters in Table \ref{tab:sfparametersNbN} \cite{charaev_magnetic-field_2019,zdravkov_reentrant_2006,ryazanov_proximity_2003}.

\begin{table}
    \centering
    \begin{tabular}{|c||c|}
        \hline
        Parameter & Value\\
        \hline \hline
       Thickness of superconducting NbN layer  $d_s$& 5 nm
 \\
    Critical Temperature  of NbN $T_c$& 14 K
\\
Normal State Resistivity of Ni$_{0.57}$Cu$_{0.43}$ $\rho_f$& 60 $\rm \mu \Omega$cm\\
Normal State Resistivity of NbN $\rho_s$& 120 $\rm \mu \Omega$cm\\
Coherence Length of NbN $\xi_s$& 4.1 nm\\
Coherence Length of Ni$_{0.57}$Cu$_{0.43}$ $\xi_f$& 7.6 nm\\
Exchange potential of Ni$_{0.57}$Cu$_{0.43}$ $E_{xc}$& 0.56 eV\\
    \hline
    \end{tabular}
    \caption{Experimentally measured parameters of Ni$_{0.57}$Cu$_{0.43}$ and NbN layer employed in our calculations \cite{charaev_magnetic-field_2019,zdravkov_reentrant_2006,ryazanov_proximity_2003}.}
    \label{tab:sfparametersNbN}
\end{table}

\bibliography{Bibliography.bib}

\end{document}